\documentclass[cameraready]{Interspeech}

\title{From Read Speech to Spoken Digits: A Task-Specific Evaluation of Speech Privacy With Informed Attackers}

\author[affiliation={1,2}, orcid=0009-0000-6687-7559, correspondingauthor]{Jule}{Pohlhausen}
\author[affiliation={3}]{Anjana}{Rajasekhar}
\author[affiliation={3}, orcid=0000-0003-2994-2336]{Anna}{Leschanowsky}
\author[affiliation={1}, orcid=0000-0002-0595-0262]{Joerg}{Bitzer}

\address{
    $^1$ Institute of Hearing Technology and Audiology, Jade University of Applied Sciences, Oldenburg, Germany\\
    $^2$ Dept. of Medical Physics and Acoustics, Carl von Ossietzky Universität Oldenburg, Germany \\
    $^3$ Fraunhofer Institute for Integrated Circuits (IIS), Erlangen, Germany
}

\email{jule.pohlhausen@jade-hs.de}


\usepackage{csquotes}
\usepackage{subcaption}
\usepackage{stfloats}
\usepackage{float}

\begin{document}

\maketitle

\begin{abstract}
Protecting speech privacy in real-life audio recordings is a growing concern.
This contribution evaluates the effectiveness of three obfuscation techniques in protecting linguistic speech content, using digit recognition as a task-specific and practically motivated evaluation scenario. 
As a first baseline, a general-purpose speech recognition model and a digit-specific classifier were applied as informed attackers to recognise both single digits and concatenated digit sequences.
Our experimental results demonstrate significant differences in recognition performance across digit modality, speech rate, and attack model.
These findings emphasize the need for more comprehensive and application-oriented evaluation methods to ensure speech privacy.
\end{abstract}

\section{Introduction}
In many areas of everyday life, audio recordings have become an integral part of data collection and automated analysis. Examples include monitoring in smart homes~\cite{hollosi2010voice} or elder care facilities~\cite{yang2025real}, acoustic surveillance systems \cite{valenzise2007scream}, and the analysis of social behaviour \cite{mehl2017ear}.
In particular, speech recordings constitute a rich source of personal, sensitive information, encompassing named entities (persons, dates, locations, organizations, etc.), medical details, political opinions, and paralinguistic attributes such as the speaker's gender or health status \cite{nautsch2019speech_privacy, williams2021revisiting}. 
Consequently, voice recordings and speech data are considered personal data under the European general data protection regulation (GDPR) and require proper protection from misuse \cite{EUGDPR}.

A common attack scenario assumes that a potential attacker has unauthorised access to audio recordings and seeks to exploit sensitive information. 
While the VoicePrivacy Challenges (VPC) \cite{voicepriv2020, Panariello2024VPC} have promoted the development of safeguards targeting speaker-related voice attributes, this contribution focuses on content-related sensitive information as a complementary direction. 
With the increasing power and availability of automatic speech recognition (ASR) systems, the risk of sensitive information being exploited has grown.
To mitigate this threat and ensure the protection of linguistic speech content, various privacy-preserving techniques have been proposed, including obfuscation, content masking, encryption, adversarial learning, and distributed learning \cite{williams2023itg, ahmed2020preech, backstrom2025privacy, nautsch2019preserving}.
Apparently, the concealment of linguistic content must be achieved while maintaining the utility of the signal for downstream tasks, resulting in a trade-off between privacy and utility.
For instance, obfuscation techniques with low computational complexity apply temporal and spectral smoothing \cite{Pohlhausen2025csl}, resampling \cite{pohlhausen2025resample}, or sound shredding \cite{kumar2015sound} to conceal linguistic content and simultaneously retain sufficient utility for tasks such as conversation analysis or audio-based activity recognition. 
Such light-weighted methods are particularly relevant for resource-constrained edge devices, such as smart home sensors or wearable recording systems, where privacy-preserving processing must be performed locally prior to any transmission or cloud-based analysis.

Most evaluations of ASR performance have relied on read speech and the word error rate (WER) as the primary evaluation metric \cite{librispeech, Alharbi2021asr, Panariello2024VPC, Pohlhausen2025csl}. However, the WER treats all words as equally important, whereas in practice only a small subset of words carries sensitive information. This motivates a shift towards task-specific evaluation, where the attack model is tailored to a specific category of sensitive information rather than general speech content.
Digits represent a particularly well-suited target for such an evaluation for several reasons. First, numerical information such as phone numbers, bank account numbers, credit card numbers, and transaction authentication numbers (TAN) is among the most directly exploitable sensitive information in speech recordings, with immediate consequences for financial security and personal privacy. Second, digits form a closed, well-defined vocabulary of ten classes, enabling controlled evaluation with clearly interpretable error rates. Third, unlike read speech, digit sequences impose a strict ordering constraint, where a single recognition error may render the entire sequence unusable, making the evaluation particularly sensitive to obfuscation effectiveness. 

The proposed evaluation assumes that a potential attacker successfully identified segments containing digit sequences in (long-term) recordings -- for instance, using voice activity detection (VAD) or keyword spotting -- and subsequently applies either a general-purpose ASR model or a digit-specific classifier for recognition. 
With accuracies exceeding 95~\% for recognising isolated digits with various architectures \cite{becker2024audiomnist, sharan2020spoken, tripathi2022sub}, the recognition of digit sequences can be addressed by first segmenting the sequence into individual digits and subsequently applying single digit recognition.
Due to data availability, the evaluation is restricted to digits from \textit{zero} to \textit{nine}, but serves as a foundation for further research that might consider multi-digit numbers, dialects, and disfluencies.
Further, the considered optimal attack conditions represent the strongest privacy concern.
To the best of the authors' knowledge, this work presents the first evaluation of the robustness of current obfuscation techniques against targeted digit recognition attacks with informed models.
In this informed scenario, the attacker has complete knowledge of the obfuscation technique and its parameters, and trains or finetunes models on obfuscated data \cite{Srivastava2020voice_conversion_attacks}.
The remainder systematically analyses the impact of digit modality, speech rate, and attack model choice on the digit recognition performance.

\vspace{-5pt}
\section{Experimental setup}\label{sec:setup}
Experiments were performed on two Nvidia RTX 3090 Ti GPU and one Intel Alder Lake CPU, based on the open-source speech-processing toolkit SpeechBrain \cite{speechbrain}.
The experimental framework is publicly available\footnote{Code available at \url{https://github.com/ol-MEGA/ppca.git}} \cite{Pohlhausen2025csl}, including scripts for training, inference, and evaluation.

\begin{table*}[!t]
    \caption{Total WER in \% on LibriSpeech test-clean \cite{Panayotov2025librispeech}: Tested with informed ASR models for temporal smoothing and resampling, while the original ASR model was applied for shredding.}
    \label{tab:wer}
    \centering
    \begin{tabular}{c|cccc|ccc|ccc}
    \toprule
    \textbf{Original} & \multicolumn{4}{c|}{\textbf{Temporal smoothing} \cite{Pohlhausen2025csl}} &\multicolumn{3}{c|}{\textbf{Resampling} \cite{pohlhausen2025resample}} &\multicolumn{3}{c}{\textbf{Shredding}} \\
    80 Mel \cite{Pohlhausen2025csl} & 125 ms & 250 ms & 375 ms & 500 ms & 800 Hz & 500 Hz & 320 Hz  & 500 ms & 200 ms & 100 ms\\
    \midrule
     2.34 &   5.97 &  32.90 &  78.43 &  102.76 &  27.55 &  52.15 & 86.91 & 102.60 & 125.30 & 153.85 \\
    \bottomrule
    \end{tabular}
\end{table*}

\subsection{Obfuscation techniques}
Three methods based on signal processing were evaluated for their ability to protect linguistic content with low computational complexity:
(1) Temporal smoothing with subsampling reduces the temporal resolution of the signal \cite{bitzer2016privacy, Pohlhausen2025csl}.
This evaluation considered smoothing times $\tau = 125$, 250, 375, and \SI{500}{\milli\second} with corresponding subsampling factors.
(2) Resampling restricts audio signals to low-frequency content \cite{pohlhausen2025resample}.
We evaluated downsampling rates of 800, 500, and \SI{320}{\hertz}.
The test signals were resampled using \texttt{torchaudio} 
with anti-aliasing filtering 
\cite{torchaudio2023}.
(3) The shredding approach \cite{kumar2015sound} segments the audio signal into fixed-length blocks, randomises their order, and concatenates the shuffled blocks to reconstruct an audio signal. 
In this evaluation, fixed block lengths of 100, 200, and \SI{500}{\milli\second} have been used.

\subsection{Attack models for digit recognition}
Our evaluation differentiates between ASR-based and digit-specific attack models.
A state-of-the-art general-purpose
ASR system\footnote{\url{https://huggingface.co/speechbrain/asr-transformer-transformerlm-librispeech}} was applied, which relies on 80-dimensional log Mel filterbank energies computed on time segments with a window size of \SI{25}{\milli\second} and a hop size of \SI{10}{\milli\second}.
Since we assume informed attackers, the transformer-based ASR model was finetuned on the LibriSpeech train-clean-360 dataset \cite{Panayotov2025librispeech} for each parameter of the first two obfuscation techniques. For shredding, the original ASR model was applied, since an informed attacker would potentially try to rearrange the fixed-length blocks to reconstruct the original signal. 
For all utilised obfuscation techniques, Table~\ref{tab:wer} summarises the total WER on LibriSpeech test-clean \cite{Panayotov2025librispeech} as a general baseline for protecting read speech.

Due to the unavailability of previous approaches to classify single digits explicitly, as the code was outdated or not publicly available \cite{becker2024audiomnist, sharan2020spoken, tripathi2022sub}, a deep neural network (DNN) with low complexity was trained \cite{kaggle2022}. 
This DNN operates on the mean and variance of 13 Mel-frequency cepstral coefficients (MFCC), aggregated over time. 
Its architecture consists of three fully connected layers, each containing 128 neurons with ReLU activation, followed by a softmax output layer with
ten classes. With only 37770 learnable parameters, the model was trained with a batch size of 32 using the Adam optimiser with sparse categorical cross-entropy as the loss function, converging after ten epochs.
The DNN-based approach is intended to be efficient under optimal attack conditions and straightforward to implement$^1$, thereby reducing the burden on potential attackers.

\subsection{Datasets}
The evaluation utilises two well-established datasets in English: AudioMNIST \cite{becker2024audiomnist} and Google Speech Commands (GSC) digits (v0.0.1) \cite{warden2018speech}. Table~\ref{tab:data} summarises general statistics for both datasets.
AudioMNIST contains 50 recordings per speaker and digit, while the GSC digits provide a more unbalanced distribution, with only 164 out of 1847 speakers having recorded all ten digits. 
The recording setup for AudioMNIST was more controlled, resulting in cleaner recordings but less speaker variation compared to the GSC digits dataset. 

In realistic scenarios, sensitive numerical information often appears as digit sequences rather than isolated digits, such as phone or bank account numbers. To simulate this, ten digits spoken by the same speaker were concatenated into a single sequence, inspired by the typical length of phone numbers \cite{phone2026}. For concatenation, we selected AudioMNIST as it provides a consistent number of recordings per speaker and digit. By varying the silence duration inserted between each digit -- specifically 100, 200, and \SI{500}{\milli\second} -- we effectively controlled the speech rate, without altering the speech rate of the original digits.
One hypothesis is that these concatenated sequences possibly provide contextual information, which improves the ASR-based digit recognition. Moreover, we expect that a higher silence duration, corresponding to a lower speech rate, will improve digit recognition for temporal smoothing, as the increased time gap between digits prevents smearing of consecutive digits.
Since the DNN is designed to process single digits, the obfuscated digit sequences were segmented based on the oracle signal boundaries and classified separately.
In practice, an attacker would need to estimate these boundaries using VAD or forced alignment; however, oracle boundaries were assumed here to provide an upper bound on recognition performance under optimal attack conditions, isolating the effect of obfuscation from segmentation errors.

\begin{table}[h]
  \caption{Statistics of AudioMNIST \cite{becker2024audiomnist} and GSC digits \cite{warden2018speech}.}
  \label{tab:data}
  \centering
  \begin{tabular}{lrr}
    \toprule
    \textbf{Dataset} & \textbf{AudioMNIST} & \textbf{GSC} \\
    \midrule
    Nr. of speakers & 60 & 1847 \\
    Total recordings & 30000 & 23666 \\
    Test recordings & 6000 & 2552 \\
    Min. - Max. signal dur. in s & 0.29 - 1.00 & 0.37 - 1.00 \\
    Median signal dur. in s & 0.63 & 1.00 \\
    \bottomrule
  \end{tabular}
  \vspace{-0.17cm}
\end{table}

\begin{figure}[!t]
    \centering
    \includegraphics[width=0.9\columnwidth]{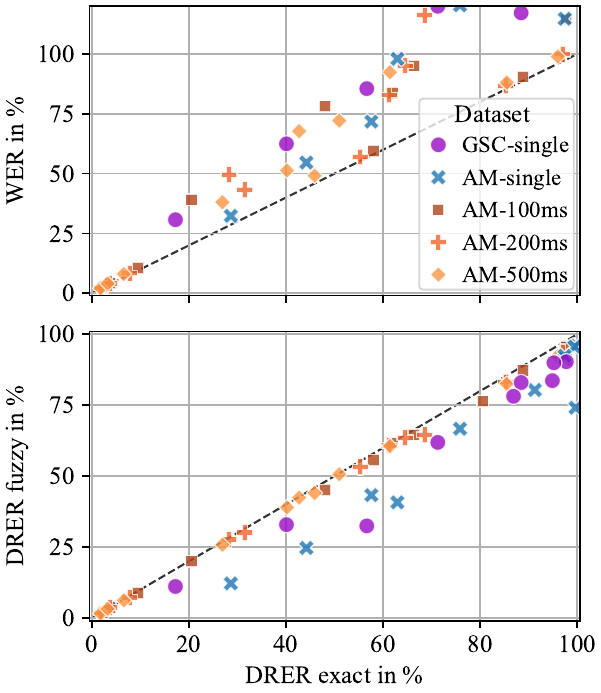}
    \vspace{-0.1cm}
    \caption{WER and DRER with exact and fuzzy matching for informed ASR models on GSC digits \cite{warden2018speech} and AudioMNIST (AM) \cite{becker2024audiomnist}, evaluated for single and concatenated digits with varying silence durations, across different obfuscation parameters.
    }
    \label{fig:asr_comp}
    \vspace{-0.35cm}
\end{figure}

\subsection{Evaluation metrics}
To facilitate comparisons with previous work \cite{Panariello2024VPC, Pohlhausen2025csl}, we determined the word error rate (WER), defined as the sum of substitutions, insertions, and deletions in the ASR transcript relative to the total number of tokens in the ground-truth transcript. 
However, for single digits, the WER may not accurately reflect the digit recognition performance due to insertions of irrelevant words in the ASR transcript.
To address this limitation, we introduce the digit recognition error rate (DRER), which compares the aligned ASR hypothesis with the reference digit only.
The DRER is evaluated using exact and fuzzy matching, where the latter considers a transcript with a Levenshtein distance \cite{levenshtein1966binary} $\leq 2$ as correct. This threshold was
motivated by a preliminary analysis of common ASR confusions, such as substituting \textit{two} with \textit{to} or \textit{too}.
To avoid inflating performance estimates due to fuzzy matching, duplicate hypotheses occurring across multiple digits,
and unrelated hypotheses 
\footnotesize{(\textit{and, here, him, I, in, it, no, non, oh, or, s, said, say, so, some, the, there, tim, you})} \normalsize
were considered incorrect.
The DRER was calculated per speaker as the number of incorrectly recognised digits divided by the total number of digits. We report the DRER averaged across speakers, along with 95~\% confidence intervals estimated via bootstrapping \cite{ferrer2024goodpracticesevaluationmachine}.

\begin{figure*}[!b]
    \centering
    \vspace{-0.1cm}
    \includegraphics[height=0.216\textheight]{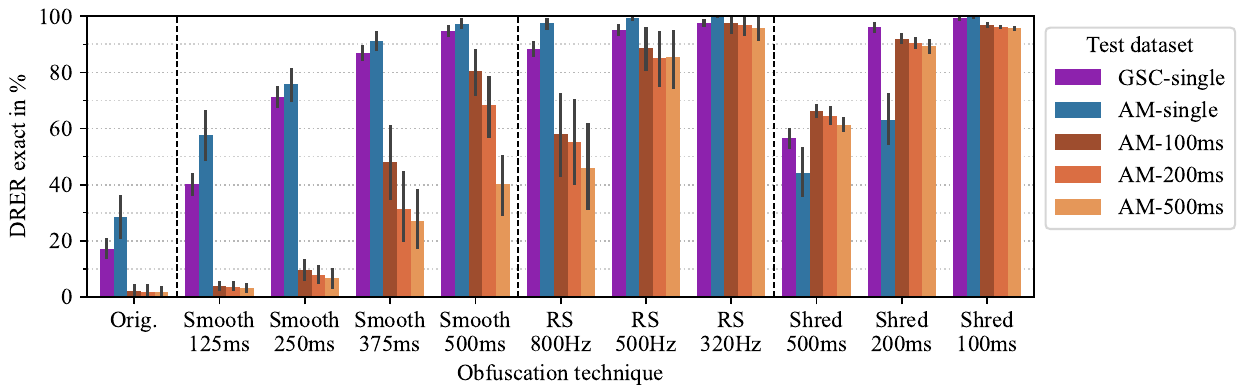}
    \caption{DRER with exact matching for informed ASR models on GSC \cite{warden2018speech} and AudioMNIST (AM) \cite{becker2024audiomnist}, both single and concatenated digits with varying silence durations. The ASR models were tested on unprotected signals (Orig.) and signals obfuscated with temporal smoothing with subsampling (Smooth), resampling (RS), and shredding (Shred). Vertical lines indicate 95~\% confidence intervals.}
    \label{fig:asr_error_concat}
\end{figure*}

\section{Results and discussion}\label{sec:results}

\subsection{ASR-based attack models}\label{sec:asr_results}
The comparison of evaluation metrics in Figure~\ref{fig:asr_comp} reveals that the DRER is consistently lower than the WER, since it ignores insertion errors. 
This difference remains stable for single (cf. purple and blue dots) and concatenated digits, as the insertion rate is similar for both modalities at higher obfuscation degrees.
However, neither metric seems ideal for digit sequences: The WER penalises inserted digits and non-digit tokens equally, while the DRER ignores insertions entirely, although they may alter the recognised sequence.
Both limitations are inherent to temporally ordered entities.
Future work will incorporate 
word-level timestamps and confidence scores to assess relevant errors.

Moreover, fuzzy matching 
reduces the DRER in Figure~\ref{fig:asr_comp} on average by 11.08~\% for single compared to 1.37~\% for concatenated digits. This difference suggests that the recognition of single digits is more prone to substitution errors, whereas ASR models seem to benefit from contextual information within a sequence of concatenated digits, making the additional tolerance of fuzzy matching less impactful. Due to space constraints, the following figures report the stricter DRER with exact matching, as 
fuzzy matching allows for more uncertainty.

Figure~\ref{fig:asr_error_concat} compares the different obfuscation techniques for both digit modalities and the influence of speech rate.
As expected, the DRER increases with higher smoothing times $\tau$, lower sampling rates, and lower shredding block lengths, reflecting the general trends on LibriSpeech test-clean (cf. Table~\ref{tab:wer}).
For the original, smoothed, and resampled single digits, the DRER in Figure~\ref{fig:asr_error_concat} is consistently lower on the GSC dataset compared to AudioMNIST, which is unexpected given that AudioMNIST is considered a cleaner dataset. However, it is likely that the presence of accents in combination with the lower number of speakers in AudioMNIST contributes to this difference.
In contrast, the shredded single digits show a lower DRER on AudioMNIST. We hypothesize that this reflects the lower median signal duration of AudioMNIST relative to the GSC digits, which are predominantly normalised to \SI{1}{\second}, as shown in Table~\ref{tab:data}. A lower signal duration indicates a higher speech ratio, which means that the main part of the digit remains unaltered during shredding and results in better recognition performance.

For temporal smoothing and resampling, the DRER is consistently lower for concatenated than for single digits. This finding can be attributed to our hypothesis that the ASR model extracts 
contextual information from digit sequences.
For digit sequences with $\tau=500$~ms, the DRER is 
up to 62.54~\% and the WER up to 51.46~\% lower than the WER on LibriSpeech test-clean in Table~\ref{tab:wer}. 
This suggests that temporal smoothing may be less effective at protecting specific entities compared to continuous read speech. Conversely, the DRER on resampled digit sequences is higher than or similar to the WER in Table~\ref{tab:wer}, possibly due to the loss of high-frequency spectral cues that are characteristic of certain English digits, such as the fricatives and plosives in \textit{six}, \textit{seven}, or \textit{three}.
Overall, these results highlight the importance of conducting realistic and practically relevant evaluations of privacy.

As expected, the DRER
declines significantly with increasing silence duration for temporal smoothing, while resampling and shredding protect more independently of silence duration. Similar to the findings on speaker identity \cite{Tomashenko2025rate}, this highlights the need for evaluating speech privacy at various speech rates.

\begin{figure*}[!t]
    \begin{subfigure}[t]{\textwidth}
        \centering
        \caption{Comparison of training and test dataset combinations}
        \includegraphics[height=0.216\textheight]{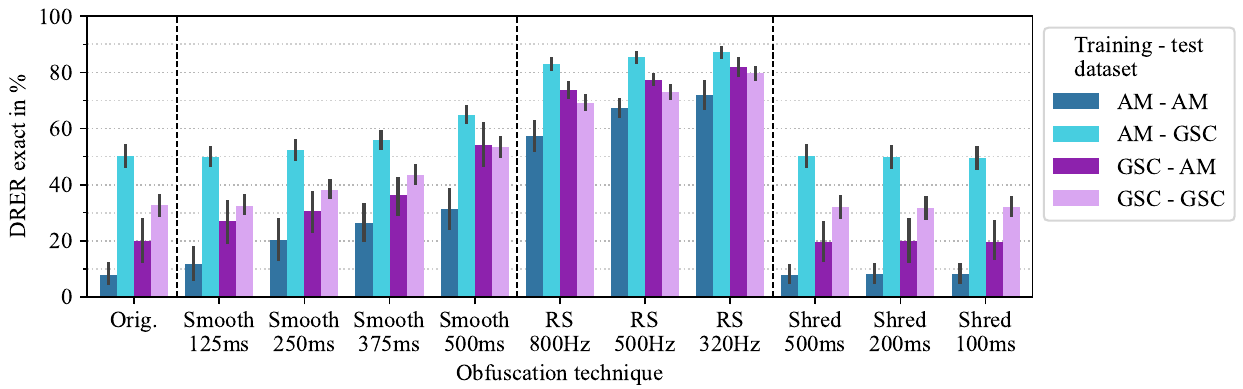}
        \label{fig:dnn_error_sets}
    \end{subfigure}
    \begin{subfigure}[t]{\textwidth}
        \centering
        \caption{Comparison of single and concatenated digits}
        \includegraphics[height=0.216\textheight]{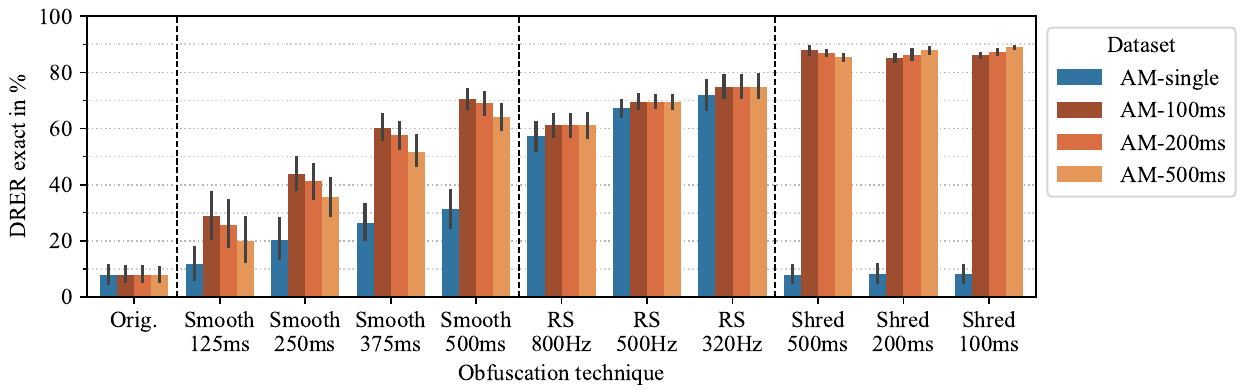}
        \label{fig:dnn_error_concat}
    \end{subfigure}
    \vspace{-0.14cm}
    \caption{DRER of DNN models trained on unprotected signals (Orig.) and with temporal smoothing with subsampling (Smooth), resampling (RS), and shredding (Shred). The DNN models were trained and tested on AudioMNIST (AM) \cite{becker2024audiomnist} and GSC digits \cite{warden2018speech} for (a) single and (b) concatenated digits with silence durations of 100, 200, and \SI{500}{\milli\second}. Vertical lines indicate 95~\% confidence intervals.}
    \label{fig:dnn}
    \vspace{-0.32cm}
\end{figure*}

\subsection{Digit-specific attack models}\label{sec:dnn_results}
The DNN-based recognition of single digits in Figure~\ref{fig:dnn_error_sets} reveals a consistent performance gap across training and test dataset combinations. The best performance is observed in the matched condition (trained and tested on AudioMNIST), while the worst results arise from the mismatched condition (AudioMNIST trained, GSC tested). 
This pattern underscores that an attacker's success is substantially contingent on access to training data drawn from a distribution similar to the target speech.
Notably, for original digits, the best combination is comparable to the performance of previous approaches \cite{becker2024audiomnist, sharan2020spoken, tripathi2022sub}.

As expected, shredding has no impact on the recognition of single digits, with the DRER remaining approximately equivalent to that observed on original, unprotected signals. This result is a direct consequence of the DNN models' input feature, consisting of the mean and variance computed over time. Since shredding only reorders temporal blocks, 
the input feature is effectively unchanged, rendering shredding entirely ineffective against this class of attacker model. Notably, this finding has broader implications for the protection of other short, precisely segmented entities with a limited set of possibilities.

In contrast to the DRER of informed ASR models shown in Figure~\ref{fig:asr_error_concat}, the DNN-based recognition of single digits deteriorates less as $\tau$ increases or the sampling rate decreases, though the general trend remains consistent. This highlights that under optimal attack conditions, a simple but task-specific attack model can pose a substantial threat to privacy.
Conversely, the DNN performed significantly worse than the ASR models on concatenated digits obfuscated with temporal smoothing or shredding, as shown in Figure~\ref{fig:dnn_error_concat}. For temporal smoothing, the DNN benefits less from an increased silence duration.
As the DNN-based DRER approaches 90~\%, corresponding to chance-level performance, the DNN models lose discriminative capability when evaluated on shredded digit sequences.

\section{Conclusions}\label{sec:con}
This article analysed the recognition of single and concatenated digits against informed attacker models. 
The superior performance of the task-specific DNN on single digits highlights that even simple attack models pose a substantial privacy threat under optimal conditions, underscoring the importance of considering a diverse range of informed attack strategies and attacker models in privacy evaluations.
In contrast, for the more realistic scenario of digit sequences, the ASR models outperformed the DNN for most obfuscation parameters, suggesting that contextual information plays a critical role in successful attacks.
Overall, experimental results demonstrate that task-specific metrics provide a more meaningful and practically relevant assessment of speech privacy than the WER on read speech alone.

\ifcameraready
\section{Acknowledgements}
This work was supported by the Graduation program of Jade University of Applied Sciences (Jade2Pro 2.0).
\fi

\section{Generative AI Use Disclosure}
During the manuscript preparation, the authors used AI-assisted writing tools in order to improve language and readability. After using these tools, the authors reviewed and edited the content as required and take full responsibility for the content of the publication.

\bibliographystyle{IEEEtran}
\bibliography{sapstrings, mybib}

\end{document}